\documentclass[letterpaper,english,aps, prl, reprint, superscriptaddress]{revtex4-1}
\usepackage[T1]{fontenc}
\usepackage[latin9]{inputenc}
\usepackage{amsmath}
\usepackage{amssymb}
\usepackage{bbold}
\usepackage{graphicx}

\makeatletter

\pdfpageheight\paperheight
\pdfpagewidth\paperwidth

\PassOptionsToPackage{position=top}{subfig}
\usepackage{hyperref}
\hypersetup{
	colorlinks = true,
	allcolors = blue
}

\usepackage{times}

\makeatother

\usepackage{babel}
\begin{document}

\title{Non-Hermitian butterfly spectra in a family of quasiperiodic lattices}

\author{Li Wang}
\email{liwangiphy@sxu.edu.cn}
\affiliation{Institute of Theoretical Physics, State Key Laboratory of Quantum Optics and Quantum Optics Devices, Collaborative Innovation Center of Extreme Optics, Shanxi University, Taiyuan 030006, P. R. China}

\author{Zhenbo Wang}
\affiliation{Institute of Theoretical Physics, State Key Laboratory of Quantum Optics and Quantum Optics Devices, Collaborative Innovation Center of Extreme Optics, Shanxi University, Taiyuan 030006, P. R. China}

\author{Shu Chen}
\email{schen@iphy.ac.cn }
\affiliation{Beijing National Laboratory for Condensed Matter Physics, Institute
of Physics, Chinese Academy of Sciences, Beijing 100190, China}
\affiliation{School of Physical Sciences, University of Chinese Academy of Sciences,
Beijing 100049, China }

\date{\today}

\begin{abstract}
We propose a family of exactly solvable quasiperiodic lattice models with analytical complex mobility edges, which can incorporate mosaic modulations as a straightforward generalization. By sweeping a potential tuning parameter $\delta$, we demonstrate a kind of interesting butterfly-like spectra in complex energy plane, which depicts energy-dependent extended-localized transitions sharing a common exact non-Hermitian mobility edge. Applying Avila's global theory, we are able to analytically calculate the Lyapunov exponents and determine the mobility edges exactly. For the minimal model without mosaic modulation, a compactly analytic formula for the complex mobility edges is obtained, which, together with analytical estimation of the range of complex energy spectrum, gives the true mobility edge. The non-Hermitian mobility edge in complex energy plane is further verified by numerical calculations of fractal dimension and spatial distribution of wave functions. Tuning parameters of non-Hermitian potentials, we also investigate the variations of the non-Hermitian mobility edges and the corresponding butterfly spectra, which exhibit richness of spectrum structures.
\end{abstract}

\maketitle

\textcolor{blue}{\em Introduction.}--In the past few decades, quasiperiodic lattices~\cite{AA1980,Thouless,Kohmoto,Kohmoto2008,Cai2013,Roati,Lahini} have  become a versatile platform to investigate disorder-induced localization transitions, which is one of key topics of fundamental importance in the frontiers of condensed matter physics. As is well known, the topic was originally proposed by Anderson in his seminal work in the context of electronic systems with truly random disorders~\cite{Anderson1958pr,RevModPhys.57.287}. While scaling theory~\cite{Abrahams1979prl} demonstrates that for truly random systems all single-particle eigenstates are already localized in one and two dimensions even under arbitrary small but finite disorder strength and thus there is no extra space left for localization transition to occur, quasicrystals containing the so-called determinant correlated disorders~\cite{AA1980,Thouless,Kohmoto,Kohmoto2008,Cai2013,Roati,Lahini} have been proven to be able to host various extended-localized transitions in low-dimensional systems.
Moreover, intriguing energy-dependent localization transitions have also been revealed in various low-dimensional quasiperiodic systems, for which under the same set of parameters extended and localized single-particle eigenstates can coexist and be separated by a critical energy $E_c$, known as mobility edge (ME)~\cite{Mott1987jpc}.
Inspired by the influential Aubry-Andr\'{e}-Harper (AAH) model, various generalized AAH-like models have been proposed~\cite{xiexc1988prl,xie90prb,Hiramoto,HanJ,Biddle10prl,Ganeshan2015prl} and discovered to be capable of accommodating localization transitions with mobility edges~\cite{DengX,Ganeshan2015prl,wyc20prl,LXJprl131.176401,Bloch,mosaic-exp,SciPostPhys.13.3.046,mprb23,vu224206,m23critical,SciPostPhys.12.1.027,XuZ,YaoH,LiX}.
Notably, a few of these models are exactly solvable \cite{Ganeshan2015prl,Biddle10prl,wyc20prl,LXJprl131.176401}, which is a fact highly appealing and significant for further exploration of the mobility edge physics.

In recent years, non-Hermitian physics has experienced a revival and become a renewed prosperous research field. Growing attention has been paid to the interplay of non-Hermiticity and quasiperiodicity \cite{Satija,JiangH,Longhi2019,LiuYX2021a,Liuyx2021,PhysRevB.101.174205,HuHui,PhysRevB.103.214202,Chen21prb,ZengQB,PhysRevB.100.125157,PhysRevB.106.024202,CaiXM2021,Longhi2021,XuZ2021,PhysRevB.105.054204,PhysRevB.105.054201,PhysRevB.106.144208,ZhouLW,XiaX2022,Datta,Gandhi,Mishra,XueP,Datta2024}.
Many previous research efforts~\cite{Chen21prb,PhysRevB.101.174205,PhysRevB.103.214202,Mishra,Longhi2019,HuHui,Gandhi,Mishra,XueP,LiuYX2021a,Liuyx2021,Datta,ZhouLW,XiaX2022} have been devoted to parity-time ($\mathcal{PT}$) symmetric~\cite{bender98} quasiperiodical systems, for which a well established correspondence between real-complex transition in eigenenergy and extended-localized transition has been revealed.
However, for a general non-Hermitian quasiperiodical system, the spectrum is usually complex, and no obligate relation between the change of spectrum structure and localization transition exits.  Although some recent works have demonstrated the existence of complex mobility edge in various non-Hermitian  quasiperiodical systems \cite{PhysRevB.106.144208,ZhouLW}, analytical results of complex mobility edges are rare and thus are particularly important for the broadening of the concept of MEs from real to the complex plane.

In this work, we propose a family of non-Hermitian quasiperiodic models with a compact analytical expression of complex mobility edges, which can further incorporate mosaic
modulations as a straightforward generalization.
Utilizing Avila's global theory, we are able to calculate the Lyapunov exponent $\Gamma(E)$ analytically and get a uniform formula of non-Hermitian mobility edge (NHME). Thus, an accurate characterization of the NHME can be implemented and its intersection with the physical spectrum produces the true mobility edge.
By varying a potential parameter $\delta$, we obtain a kind of intriguing butterfly-like spectra in complex energy plane. For a typical system, we showcase that the extended and localized states distribute on the body  and wings of butterfly, respectively, separated by the NHMEs.  
Varying the non-Hermiticity parameter $\gamma$ can lead to the change of spectrum structure and NHME. We also show the deformation of the non-Hermitian butterfly spectrum for systems with various parameters.

\begin{figure}[tp]
\includegraphics[scale=0.03]{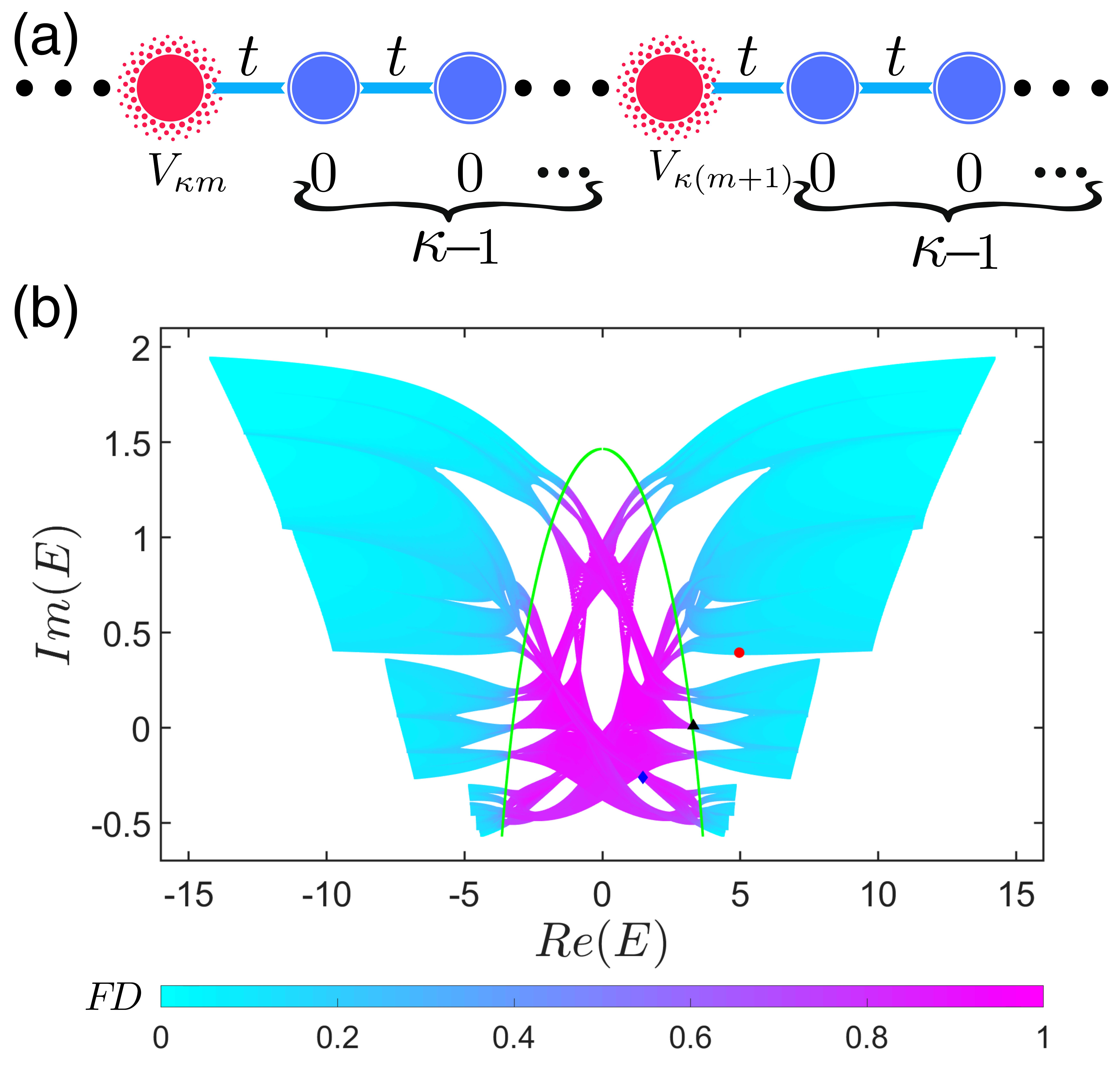}
\caption{\label{FD}(a) Sketch of the family of 1D non-Hermitian quasiperiodic lattice model. The solid  blue lines denote the homogenous hopping $t$. The red circles denote lattice sites with non-Hermitian quasiperiodic potential $V_{\kappa m}$ with $m$ being an integer, while the other circles in between denote lattice sites with zero potential. (b) The non-Hermitian butterfly spectrum of the minimal model with $\kappa=1$. Fractal dimension (FD) of each eigenstate is encoded in the color of each energy point in the spectrum. The true non-Hermitian mobility edge is denoted by a green line which is given by a comprehensive consideration of both Eq.(\ref{k1ME}) and the actual range of the model's spectrum.
Parameters: $L=987$, $\lambda=1$, $\alpha=0.5$, $\gamma=\pi/2$, $\theta=0$, $t=1$, and the modulation parameter $\delta$ varies from $-7$ to $7$.}
\label{Fig01}
\end{figure}

\textcolor{blue}{\em Model and non-Hermitian butterfly spectrum.}--We propose a family of generic non-Hermitian quasiperiodic models which are described in a unified manner by the following eigenvalue equation,
\begin{equation}
t \left ( \phi_{j-1}+\phi_{j+1} \right) + V_j \phi_j =E \phi_j,
\label{vavbEqa}
\end{equation}
where $j$ is the index of lattice site, and $t$ is the nearest-neighbour hopping amplitude. The core feature of the model Eq.(\ref{vavbEqa}) then is the non-Hermitian quasiperiodic mosaic~\cite{wyc20prl} on-site potential
with
\begin{equation}
V_j =\left\{
                \begin{array}{cc}
                 \frac{\lambda e^{i \gamma}\cos(2\pi j b+\theta)+\delta}{1-\alpha \cos(2\pi j b+\theta )}, & j=m \kappa, \\
                  0, & \mathrm{otherwise}, \\
                \end{array}
              \right. \label{vj}
\end{equation}
in which $\kappa$ is a positive integer and $m=1,2,...,N$. Apparently, the quasiperiodic potential occurs periodically with period $\kappa$, which is pictorially shown in Fig.\ref{Fig01}(a). $N$ can be seen as the number of quasicells, therefore the lattice size of the model is $L=N\kappa$. So $\kappa=1$ is for the usual quasiperiodic lattice while each $\kappa \ge 2$ corresponds to a mosaic quasiperiodic lattice~\cite{wyc20prl,LiuYX2021a}. Here the quasiperiodic on-site potential is controlled by two modulation parameters $\lambda$, $\delta$ and a deformation parameter $\alpha$. The parameter $b$ is an irrational number which is responsible for the quasiperiodicity of the on-site potential. To be concrete and without loss of generality, in this work we choose $b=(\sqrt{5}-1)/2$. The parameter $\gamma$ is a phase angle dictating the non-Hermitian nature of the on-site quasiperiodic potential and $\theta$ denotes a phase offset. Obviously, a non-Hermitian potential in this form does not respect parity-time ($\mathcal{PT}$) symmetry which is otherwise a key ingredient of previous works addressing non-Hermitian localization transitions~\cite{Chen21prb,PhysRevB.101.174205,PhysRevB.103.214202,Mishra,Longhi2019,HuHui,Gandhi,Mishra,XueP,LiuYX2021a,Liuyx2021,Datta,ZhouLW,XiaX2022}. For convenience, we shall set $t=1$ as the energy unit in the following calculation.


In this work, we shall study the general non-Hermitian case with $\kappa \ge 1$ and $\alpha \in (-1,1)$ in the presence of both $\lambda$ and $\delta$ terms. By applying Avila's global theory, we can derive NHMEs analytically by calculating the Lyapunov exponents for the general case with $\kappa \ge 1$.
We shall prove that the model has exact NHMEs separating localized states and extended states. However, to facilitate our discussion we focus on the minimal model with $\kappa=1$ and then showcase examples with $\kappa \ge 2$.
For the minimal model, a compactly analytical formula for the non-Hermitian mobility edges (NHMEs) can be obtained,
\begin{equation}
\left[\alpha Re(E)+\lambda \cos\gamma\right]^2+\frac{\left [\alpha Im(E)+\lambda \sin\gamma\right]^2}{1-\alpha^2}=4t^2,
\label{k1ME}
\end{equation}
where $Re(E)$ and $Im(E)$ are respectively the real and imaginary parts of $E$. Eq.(\ref{k1ME}) is our key result. For the general case with $\gamma \neq n \pi$,  Eq.(\ref{k1ME}) indicates that the ME takes a complex value, which is irrelevant to the parameter $\delta$. When  $\gamma=0$,  the potential $V_j$ is real, and the model reduces to the generalized Ganeshan-Pixley-Das Sarma (GPD) model~\cite{Ganeshan2015prl,prb10817,Cai_2023}. Since $E$ takes a real value, Eq.(\ref{k1ME}) is simplified to $\alpha Re(E)+\lambda \cos\gamma = \pm 2t$, consistent with the results of generalized GPD model.

Before proceeding with a rigorous proof of Eq.(\ref{k1ME}), we first conduct a numerical verification to gain an intuitive understanding.
In Fig.~\ref{Fig01}(b), by implementing numerical calculations we display in the complex plane the energy spectrum of Eq.(\ref{vavbEqa}) with the color encoding the fractal dimension (FD) of the corresponding eigenstate.
For an arbitrary normalized eigenstate $\phi$, the fractal dimension is defined as $F\!D=-\lim_{L\rightarrow \infty} \ln(\sum_j |\phi_j|^4)/\ln L$, which acts as a good indicator for distinguishing localized and extended states in that $F\!D\!\rightarrow\!0$ for localized states and $F\!D\!\rightarrow\!1$ for extended states.
As the modulation parameter $\delta$ varies, we get intriguing non-Hermitian butterfly spectra displayed in Fig.~\ref{Fig01}(b), in which the localized states and extended states are well separated. Obviously, the separation between localized and extended states is energy-dependent in the complex-energy plane, which coincides well with the green line  plotted according to the analytical  formula Eq.(\ref{k1ME}). Notably, the resultant appearance of the NHME denoted by the green line in Fig.~\ref{Fig01}(b) is actually obtained after a proper estimation of the range of the model's energy spectrum. Such an estimation in detail is given in the Supplemental Material~\cite{SM}. In Fig.~\ref{Fig01}(b), the lattice size is $L=987$, $\lambda=1$, $\alpha=0.5$, $\gamma=0.5\pi$, $\theta=0$, and the modulation parameter $\delta$ varies from $-7$ to $7$.

\begin{figure}[tbp]
\centering
\includegraphics[scale=0.33]{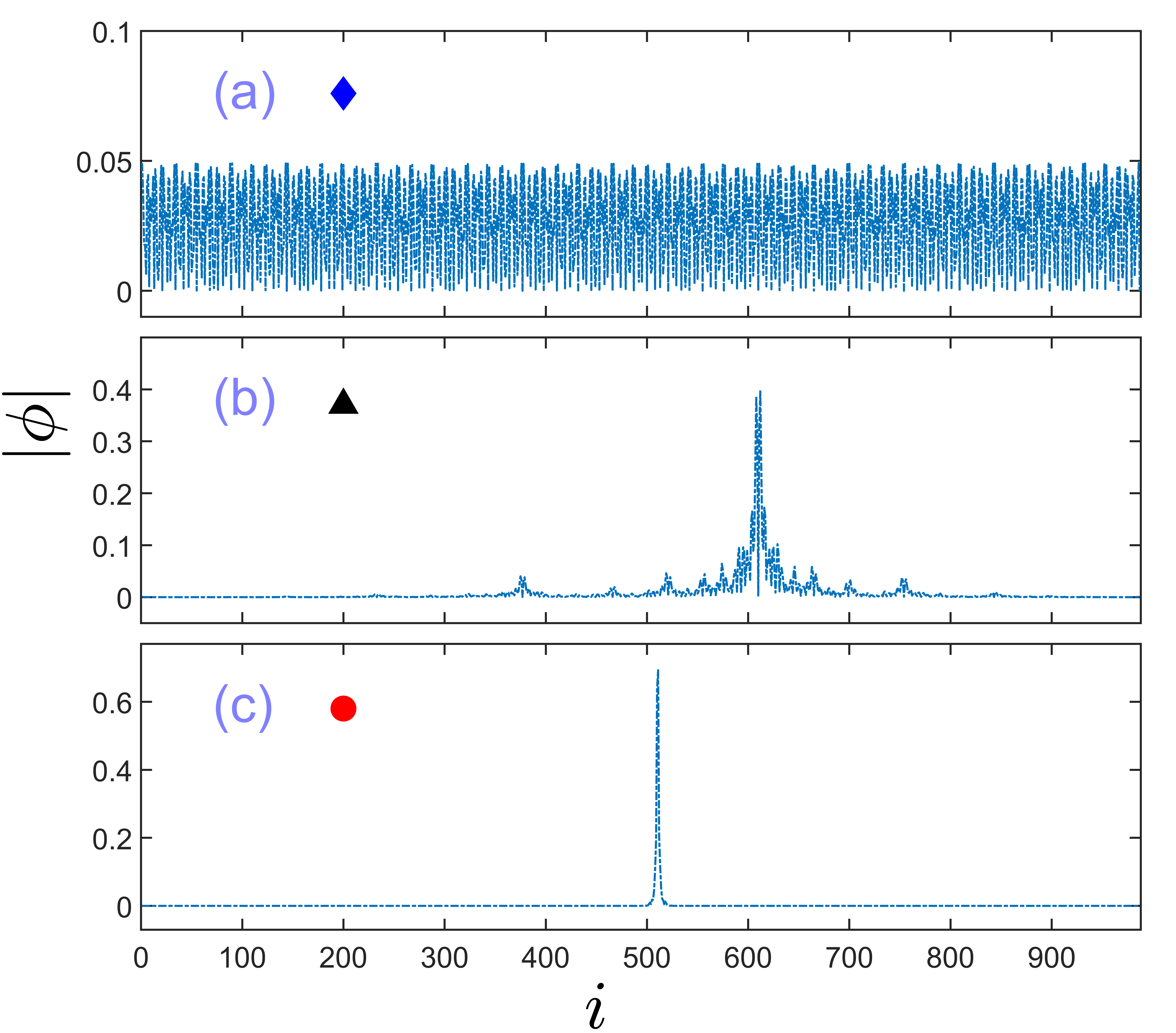}
\caption{Typical spatial distributions of eigenstates in different regions of the non-Hermitian butterfly spectrum.
(a) Extended state corresponding to eigenenergy denoted by a blue diamond in Fig.~\ref{Fig01}.
(b) Critical state corresponding to eigenenergy denoted by a black triangle in Fig.~\ref{Fig01}.
(c) Localized state corresponding to eigenenergy denoted by a red dot in Fig.~\ref{Fig01}.
Model parameters are the same as in Fig.~\ref{Fig01}.}
\label{Fig02}
\end{figure}

Representative distributions of eigenstates corresponding to different regions of the non-Hermitian butterfly spectrum in Fig.~\ref{Fig01}(b) is shown in Fig.~\ref{Fig02}. The three eigenstates are denoted respectively by blue diamond, black triangle, and red dot in the non-Hermitian butterfly spectrum. Clearly, Fig.~\ref{Fig02}(a) and Fig.~\ref{Fig02}(c) display typical extended state and localized state, respectively, while Fig.~\ref{Fig02}(b) shows critical state with multifractal structure.

\textcolor{blue}{\em Analytical derivation of NHME.}
--Analytically, the non-Hermitian mobility edge of the generic non-Hermitian quasiperiodic mosaic model in Eq.~(\ref{vavbEqa}) can be exactly derived by computing  Lyapunov exponent. For convenience, $t$ is absorbed into $\lambda$ and $E$ in the intermediate derivation process and will be restored later.

According to Avila's global theory of one-frequency analytical $S\!L(2,\mathbb{R})$ cocycle\cite{avila}, the Lyapunov exponent $\Gamma(E)$ can be calculated as
\begin{equation}
\Gamma(E)=\lim_{N\rightarrow \infty} \frac{1}{N\kappa}  \ln \left\vert \left\vert \prod\nolimits_{m=1}^{N} T_m \right\vert\right\vert,
\label{LEa}
\end{equation}
where $\|\cdot\|$ denotes the norm of matrix.
$T_m$ is the one-step transfer matrix of the Schr{\"o}dinger operator at the $m$-th quasicell, which can be explicitly written as
\begin{equation}
T_m=
\begin{pmatrix}
E-V_{\kappa m}& -1 \\
1 & 0
\end{pmatrix}
\begin{pmatrix}
E & -1 \\
1 & 0
\end{pmatrix}^{\kappa-1}
\end{equation}
with $V_{\kappa m}$ given by Eq.(\ref{vj}).

To ease the calculation of $\Gamma(E)$ according to Eq.(\ref{LEa}), one can reorganise $T_m$ as
\begin{equation*}
T_m=\frac{Y_m}{X_m},
\end{equation*}
where
\begin{eqnarray*}
X_m&=&1-\alpha \cos(2\pi b \kappa m+\theta), \\
Y_m&=&
\begin{pmatrix}
EX_m-V_{\kappa m}X_m& -X_m \\
X_m & 0
\end{pmatrix}
\begin{pmatrix}
a_{\kappa} & -a_{\kappa-1} \\
a_{\kappa-1} & -a_{\kappa-2}
\end{pmatrix}.
\end{eqnarray*}
Noting that, a convenient mathematical relation has been implemented above, as
\begin{equation*}
\begin{pmatrix}
E & -1 \\
1 & 0
\end{pmatrix}^{\kappa-1}
=\begin{pmatrix}
a_{\kappa} & -a_{\kappa-1} \\
a_{\kappa-1} & -a_{\kappa-2}
\end{pmatrix},
\end{equation*}
in which
\begin{equation}
a_{\kappa}=\frac{1}{D}\left[ \left(\frac{E+D}{2}\right)^{\kappa}-\left(\frac{E-D}{2}\right)^{\kappa}\right]
\end{equation}
with $D=\sqrt{E^2-4}$.

In this way, the Lyapunov exponent can be rewritten as
\begin{equation}
\Gamma(E)=\lim_{N\rightarrow \infty} \frac{1}{N\kappa} \left[ \ln \left\| \prod\nolimits_{m=1}^{N} Y_m \right\|-  \sum\nolimits_{m=1}^{N} \ln \left | X_m \right | \right],
\label{LEb}
\end{equation}
in which
\begin{eqnarray*}
\lim\limits_{N\rightarrow \infty} \frac{1}{N\kappa}   \sum_{m=1}^{N} \ln \left |X_m \right |
&=& \frac{1}{2\pi\kappa}\int_{0}^{2\pi} \ln (1-\alpha \cos\varphi)d \varphi \nonumber \\
&=& \frac{1}{\kappa} \ln \frac{1+\sqrt{1-\alpha^2}}{2}.
\end{eqnarray*}
With the above preparations, we can now focus on tackling the remaining part of Eq.(\ref{LEb}).
Following the typical procedure of Avila's global theory\cite{avila}, the first step is to perform an analytical continuation of the phase in $Y_m$, i.e. $\theta\rightarrow\theta+i\epsilon$.
Considering large-$\epsilon$ limit, a straightforward derivation leads to
\begin{align*}
Y_m(\epsilon)&=\frac{1}{2} e^{-i(2\pi b\kappa m+\theta)} e^{\epsilon} \nonumber \\
& \times
\begin{pmatrix}
-\chi a_{\kappa}+\alpha a_{\kappa-1} & \chi a_{\kappa-1}-\alpha a_{\kappa-2} \\
-\alpha a_{\kappa} & \alpha a_{\kappa-1}
\end{pmatrix}
+o(1),
\end{align*}
where $\chi=\alpha E+\lambda e^{i\gamma}$. This accordingly leads to
\begin{align*}
\lim_{N\rightarrow \infty} \frac{1}{N\kappa} \ln \left\| \prod\nolimits_{m=1}^{N} Y_m \right\|
=\frac{1}{\kappa} \epsilon+\frac{1}{\kappa}\ln f,
\end{align*}
in which,
\begin{align}
\label{f}
f=\mathrm{max}\left\{ \left|\frac{2\alpha a_{\kappa -1}-\chi a_{\kappa}\pm G}{4}\right | \right \},
\end{align}
with
\begin{align}
G=\sqrt{\chi^2 a_{\kappa}^2-4\alpha\chi a_{\kappa}a_{\kappa-1}+4\alpha^2 a_{\kappa} a_{\kappa-2}}.
\end{align}
Thus, we have
$\kappa \Gamma_{\epsilon}(E)=\epsilon+\ln  \frac{2 f}{1+\sqrt{1-\alpha^2}}$.
Avila's global theory~\cite{avila} shows that, $\kappa \Gamma_{\epsilon}(E)$ is a convex, piecewise linear function of $\epsilon$ with integer slopes.
This implies that
$\kappa \Gamma_{\epsilon}(E)=\mathrm{max}\{\epsilon+\ln  \frac{2 f}{1+\sqrt{1-\alpha^2}} , \kappa \Gamma_0(E)\}$.
Furthermore, Avila's global theory proves that, $E$ does not belong to the spectrum, if and only if $\Gamma_0(E)>0$, and $\Gamma_{\epsilon}(E)$ is an affine function in a neighborhood of $\epsilon=0$.
Therefore, for any $E$ lies in the spectrum, we have
\begin{align}
\Gamma(E)=\frac{1}{\kappa}\mathrm{max}\{\ln  \frac{2 f}{1+\sqrt{1-\alpha^2}} , 0\}.
\end{align}
Then NHMEs for general $\kappa$ can be determined exactly by letting $\Gamma(E)=0$.
Notably, a compact formula Eq.(\ref{k1ME}) of NHMEs for the simplest case $\kappa=1$ can be obtained in this way. NHMEs for other cases with $\kappa \ge 2$ could be found in the Supplemental Material~\cite{SM}.

\begin{figure}[tp]
\includegraphics[scale=0.14588]{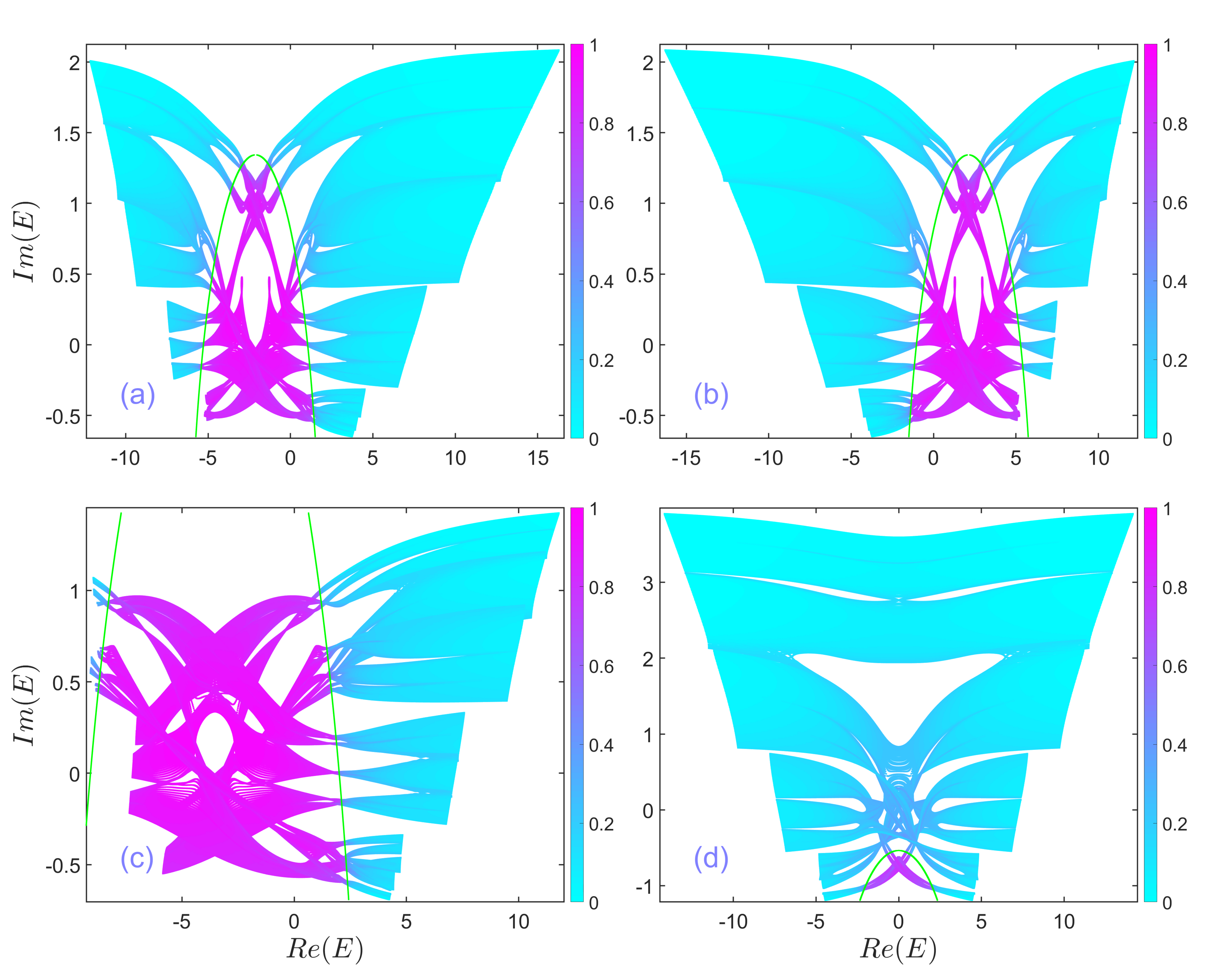}
\caption{\label{Variation}
Variations of the non-Hermitian butterfly spectrum for the minimal model with $\kappa=1$ by tuning model parameters. (a) $\gamma=\pi/4$, $\lambda=1.5$, and $\alpha=0.5$. (b) $\gamma=3\pi/4$, $\lambda=1.5$, and $\alpha=0.5$. (c) $\gamma=\pi/4$, $\lambda=1.5$, and $\alpha=0.3$. (d) $\gamma=\pi/2$, $\lambda=2$, and $\alpha=0.5$.
Fractal dimension (FD) of each eigenstate is denoted by the color of each energy point in the spectrum. The true non-Hermitian mobility edge is denoted by a green line which is given by a comprehensive consideration of both Eq.(\ref{k1ME}) and the actual range of the model's spectrum.
Other parameters are as follows, $L=987$, $t=1$, $\theta=0$, and the modulation parameter $\delta$ varies from $-7$ to $7$.}\label{Fig03}
\end{figure}

\textcolor{blue}{\em Variations of the non-Hermitian butterfly spectrum.}--The richness and exact solvability of the proposed generic non-Hermitian quasiperiodic model given in Eq.(\ref{vavbEqa}) grants us plenty of tangible freedom to introduce variations to the interesting butterfly spectrum and the exact non-Hermitian mobility edge within the spectrum.

As the concrete workhorse, the minimal model with $\kappa=1$ possesses a favorable and compact analytic formula Eq.(\ref{k1ME}) for the exact non-Hermitian mobility edge, which is simply an ellipse equation with properties that are familiar to us. Apparently, the center of the exact NHME lies at the point $(-\lambda \cos\gamma/\alpha,-\lambda \sin \gamma /\alpha)$ of the complex plane. As the non-Hermiticity parameter $\gamma$ varies from $0$ to $2\pi$, the center of the ellipse which is also the center of possibly presented extended-region in the butterfly spectrum, will run around a circle. In other words, the parameter $\gamma$ determines the orientation of the center of the ellipse, while the ratio between $\lambda$ and $\alpha$ controls the distance of the ellipse center from the origin of the complex plane.
In Fig.~\ref{Fig03}(a) and Fig.~\ref{Fig03}(b), we show the non-Hermitian butterfly spectrum and the corresponding non-Hermitian mobility edge line for the minimal model with different non-Hermiticity parameter $\gamma$, i.e., (a) $\gamma=\pi/4$ and (b) $\gamma=3\pi/4$. It is clearly shown that the orientation of the non-Hermitian mobility edge line is dependent on the non-Hermiticity parameter $\gamma$.

Moreover, it is obvious that the semi-major axis and the semi-minor axis of the ellipse are $a=|2t/\alpha|$ and $b=|2t\sqrt{1-\alpha^2}/\alpha|$, respectively. As hopping amplitude $t$ has been set to be the energy unit throughout the paper, the deformation parameter $\alpha$ is the sole parameter which can be used to monitor the size of the ellipse. Since the energy area inside the ellipse in the complex plane corresponds to extended eigenstates, changing the value of $\alpha$ may alter the portion of extended eigenstates in the whole spectra. This can be clearly illustrated by comparing Fig.~\ref{Fig03}(a) and Fig.~\ref{Fig03}(c). Compared to Fig.~\ref{Fig03}(a), the deformation parameter $\alpha$ for Fig.~\ref{Fig03}(c) decreases from $0.5$ to $0.3$, resulting in the obvious enlargement of the extended region in the exotic non-Hermitian butterfly spectra.
\begin{figure}[b]
\centering
\includegraphics[width=7.7cm]{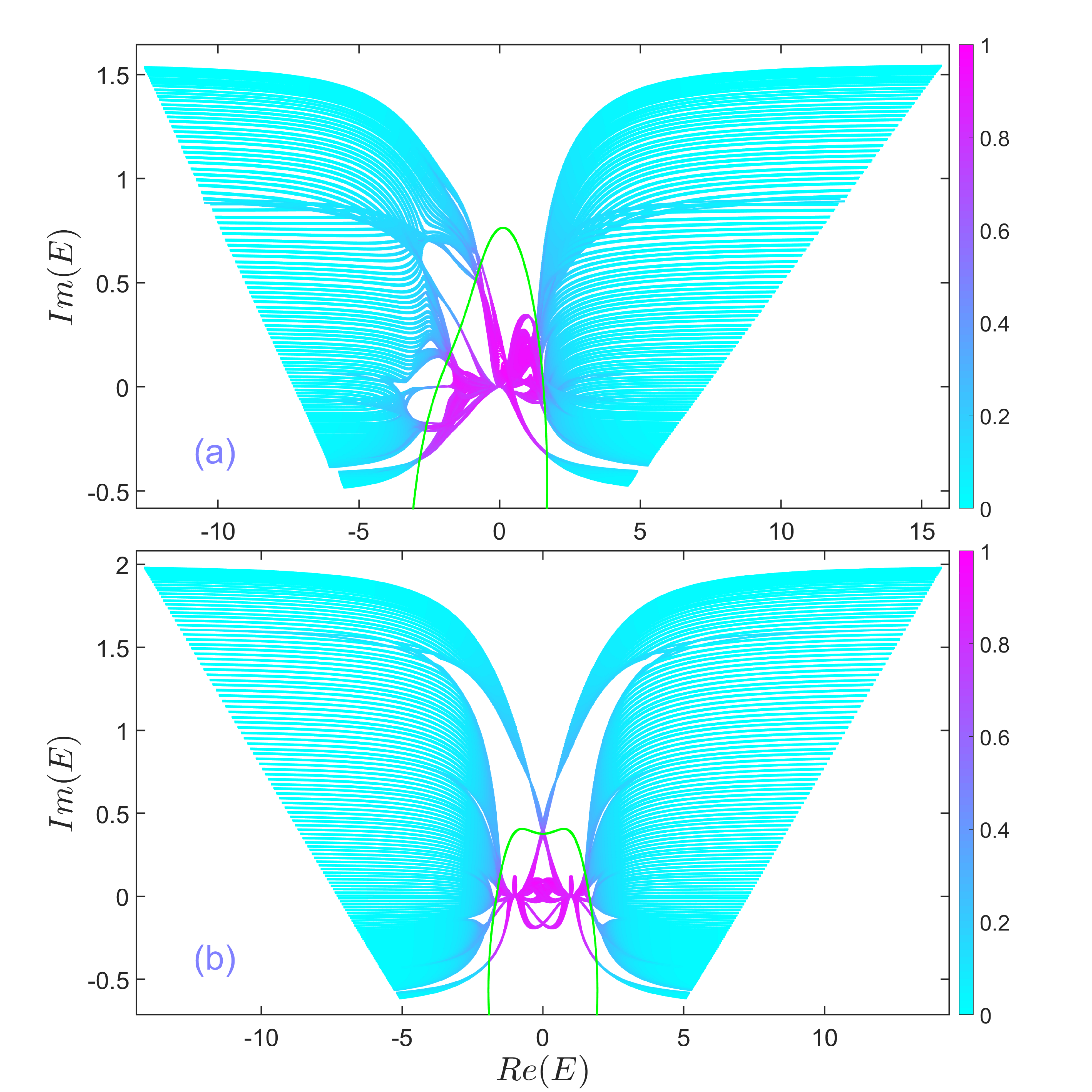}
\caption{The non-Hermitian butterfly spectrum of the generic non-Hermitian mosaic quasiperiodic model with (a) $\kappa=2$ and (b) $\kappa=3$. Fractal dimension (FD) of each eigenstate is denoted by the color of each energy point in the spectrum. The exact non-Hermitian mobility edge is denoted by a green line which is obtained by numerically solving the exact relation Eq.(\ref{MErelation23}) and considering the actual range of the model's spectrum at the same time.
Parameters: (a) $L=610$, $\lambda=1.1$, $\alpha=0.5$, $\gamma=\pi/4$, $\theta=0$, $t=1$, (b): $L=987$, $\lambda=1$, $\alpha=0.5$, $\gamma=\pi/2$, $\theta=0$, $t=1$. The modulation parameter $\delta$ varies from $-7$ to $7$.}
\label{K2K3ME}
\end{figure}

Also, knowledge obtained from investigation on AAH-like models tells that increasing the value of $\lambda$ may increase the number of localized eigenstates and decrease the number of extended eigenstates. This fact is illustrated in Fig.~\ref{Fig03}(d). For Fig.~\ref{Fig01}(b) and Fig.~\ref{Fig03}(d), all model parameters are the same except $\lambda$ increased from $1$ to $2$.
As a result, the extended region in the non-Hermitian butterfly spectrum clearly shrinks.
It is worth noting that in order to precisely modulate the interesting non-Hermitian butterfly spectrum and the non-Hermitian mobility edge, we also need to consider the structure and the actual distribution range of the system's energy spectrum. A rough analytic estimation of the range of the model's complex energy spectrum is provided in the supplemental material~\cite{SM}.

Before ending the letter, we would like to showcase the spectra and NHMEs for the cases with $\kappa=2$ and $\kappa=3$. The exact NHME for general $\kappa$ can be obtained by solving the following exact relation \cite{SM},
\begin{align}
\frac{2 f}{1+\sqrt{1-\alpha^2}}=1
\label{MErelation23}
\end{align}
with $f$ given by Eq.(\ref{f}).
As shown in Fig.~\ref{K2K3ME}(a) and Fig.~\ref{K2K3ME}(b), both spectra display intriguing butterfly-like structures and the numerical results agree well with curves of NHMEs plotted according to solving Eq.(\ref{MErelation23}) with $\kappa=2$ and $\kappa=3$, respectively.

\textcolor{blue}{\em Summary.}--In summary, we have proposed a family of exactly solvable 1D quasiperiodic lattice models with complex MEs. With the help of Avila's global theory, we derived a compactly analytical formula of NHMEs, which indicates clearly how the complex mobility edges form and are affected by modulation parameters.
Our models exhibit intriguing butterfly-like  spectra in the complex energy plane with extended and localized states separated by NHMEs.
Tuning parameters of non-Hermitian potentials leads to the change of the NHMEs and deformation of butterfly spectra, which exhibit rich structures.
Our models can be directly extended to cases incorporated mosaic modulations. Our analytical results provide a firm ground for the broadening of the concept of MEs from real to the complex plane.

\textcolor{blue}{\em Acknowledgments}--
L.W. is supported by the Fundamental Research Program of Shanxi Province, China (Grant No. 202203021211315), the National Natural Science Foundation of China (Grant Nos. 11404199, 12147215) and the Fundamental Research Program of Shanxi Province, China (Grant Nos. 1331KSC and 2015021012). S. C. is supported by  by National Key Research and Development Program of China (Grant No. 2023YFA1406704), the NSFC under Grants No. 12174436 and
No. T2121001 and the Strategic Priority Research Program of Chinese Academy of Sciences under Grant No. XDB33000000.


%


\renewcommand{\thesection}{S-\arabic{section}}
\setcounter{section}{0}  
\renewcommand{\theequation}{S\arabic{equation}}
\setcounter{equation}{0} 
\renewcommand{\thefigure}{S\arabic{figure}}
\setcounter{figure}{0}   
\renewcommand{\thetable}{S\Roman{table}}
\setcounter{table}{0}    
\onecolumngrid \flushbottom

\newpage

\begin{center}
\large \textbf{Supplementary Material}
\end{center}

This Supplemental Material provides additional information for the main text. In Sec.~\ref{spectrum}, we demonstrate more details of the energy spectrum characteristics of the general non-Hermitian quasiperiodic model described by Eq.(\ref{vavbEqa}).
In order to obtain the actual non-Hermitian mobility edges, one should also take into account the actual range of the model's energy spectrum in addition to the exact non-Hermitian mobility edges obtained through the analytical calculation of Lyapnov exponents.
Thus, in Sec.~\ref{RangofE}, a rough estimation of the model's practical energy spectrum is given and numerical verification is also provided therein.
As mentioned in the main text, the analytical derivation of the exact NHME for the generic non-Hermitian quasiperiodic model also apply to cases with larger $\kappa$s.
In Sec.~\ref{largeK}, we present the
non-Hermitian Butterfly Spectra and the corresponding exact non-Hermitian mobility edges for cases with larger $\kappa$s other than the simplest case with $\kappa=1$ which has been intensively discussed in the main text.

\section{The spectrum of the generic non-Hermitian model}\label{spectrum}
\begin{figure}[!h]
\includegraphics[scale=0.5]{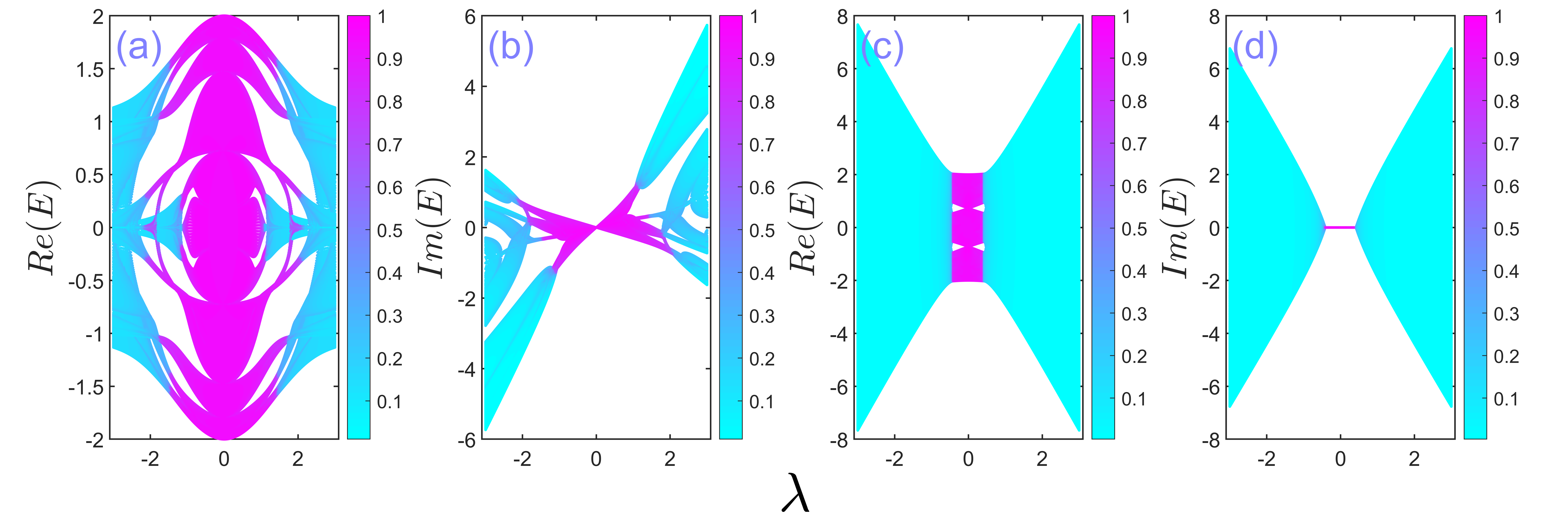}
\caption{Energy spectrum of non-Hermitian models as a function of the potential parameter $\lambda$. (a) and (b) show respectively the real parts and the imaginary parts of the eigenvalues for the generic non-Hermitian model in Eq.(\ref{vavbEqa}) with $\kappa=1$, $L=987$, $t=1$, $\delta=0$, $\theta=0$, $\alpha=0.5$, and $\gamma=\pi/2$.
(c) and (d) display respectively the real parts and the imaginary parts of the eigenvalues for the non-Hermitian $\mathcal{PT}$-symmetric AAH model in Eq.(\ref{NHAAmodel}) with $L=987$, $t=1$, $h=\pi/2$.
The color of each eigenenergy point denotes the fractal dimension of the corresponding eigenstate. }
\label{FigS1}
\end{figure}

For the model described by Eq.(\ref{vavbEqa}), the non-Hermitian physics is mainly governed by a general exponential term $e^{i\gamma}$. Apparently, this Hamiltonian does not possess an explicit $\mathcal{PT}$ symmetry.
Therefore, most of its eigenvalues are genuine complex with non-zero imaginary parts, as is shown in Fig.\ref{FigS1}(b) corresponding to the simplest case with $\kappa=1$. Unlike non-Hermitian quasiperiodic systems with $\mathcal{PT}$ symmetry, real-complex transition is generally absent from the spectrum for the model Eq.(\ref{vavbEqa}). No obligate connection between the change of energy spectrum  structure and extended-localized transition exists. This is in stark contrast with $\mathcal{PT}$-symmetric non-Hermitian model, for which there exists well-established correspondence between real-complex transition in energy and extended-localized transition. To make an intuitive comparison, we also calculate the energy spectrum of a brief and elegant $\mathcal{PT}$-symmetric AAH model, which is described by an eigenvalue equation as follows,
\begin{equation}
t \left ( \phi_{j-1}+\phi_{j+1} \right) + \lambda \cos(2\pi j b+i h) \phi_j =E \phi_j,
\label{NHAAmodel}
\end{equation}
where $i h$ denotes a complex phase factor dictating the non-Hermiticity of the $\mathcal{PT}$-symmetric AAH model. $t$ denotes the hopping amplitude, $j$ is the lattice site index, $\lambda$ describe the stength of the quasiperiodic potential, and $b$ an irrational number responsible for the quasiperiodicity of the lattice.
From Fig.\ref{FigS1}(c) and Fig.\ref{FigS1}(d), one can clearly see that extended-localized transitions are well coincident with real-complex transitions in the spectrum. However, this appealing connection is absent in Fig.\ref{FigS1}(b) for a generic non-Hermitian quasiperiodic model like Eq.(\ref{vavbEqa}).
Therefore, it is no longer possible to describe the fundamental extended-localized transitions solely exploiting the real parts of eigenenergies.

However, on the other hand, one can clearly see that energy-dependent localization transitions indeed occur as $\lambda$ varies. So a natural question arises: How to appropriately describe the localization transitions and also the possible mobility edges for this family of generic non-Hermitian quasiperiodic models.
This precisely constitutes the basic motivation of this work.
Since it turns out that only utilizing the real parts of eigenvalues is not adequate, it may imply that one should also consider the imaginary parts of eigenvalues and tackle this problem in the full complex energy plane. We have adopted this idea in the main text.

\begin{figure}[!h]
\includegraphics[scale=0.5]{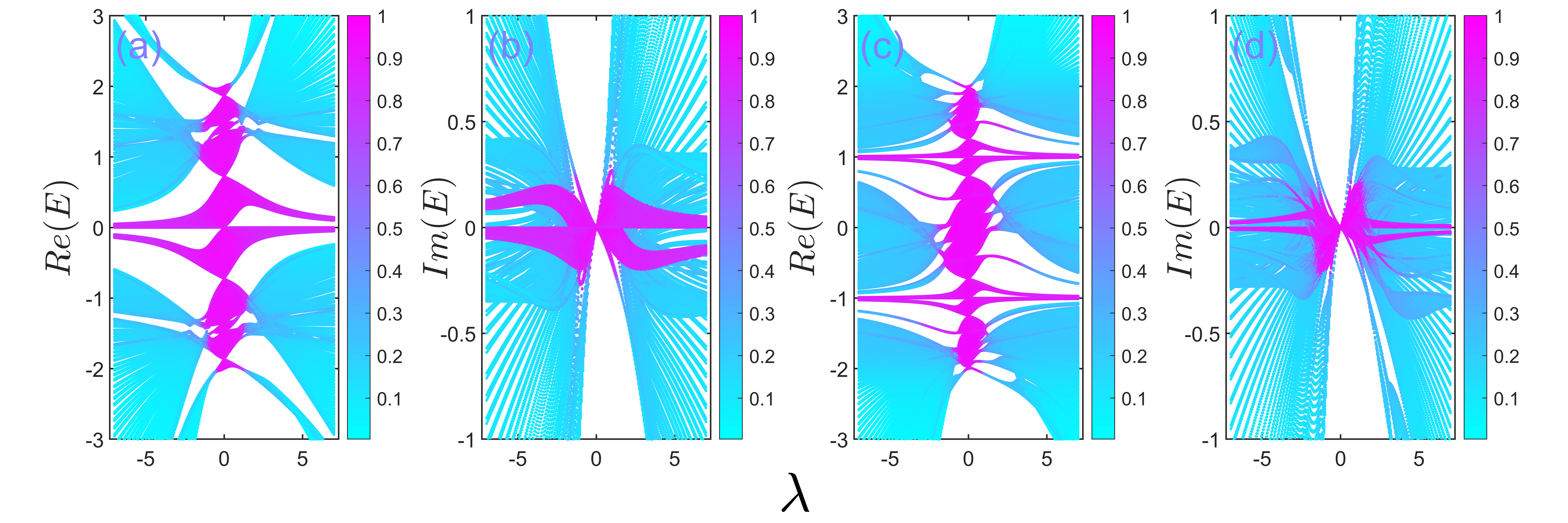}
\caption{The example energy spectrum of the generic non-Hermitian model with $\kappa \ge 2$ as a function of the potential parameter $\lambda$. (a) and (b) show respectively the real parts and the imaginary parts of the eigenvalues for the generic non-Hermitian model in Eq.(\ref{vavbEqa}) with $\kappa=2$, $L=610$, $t=1$, $\delta=0$, $\theta=0$, $\alpha=0.5$, and $\gamma=\pi/4$.
(c) and (d) display respectively the real parts and the imaginary parts of the eigenvalues for the generic non-Hermitian model in Eq.(\ref{vavbEqa}) with $\kappa=3$, $L=987$. And other parameters are the same as the case of $\kappa=2$.
The color of each eigenenergy point denotes the fractal dimension of the corresponding eigenstate. }
\label{FigS2}
\end{figure}

In Fig.\ref{FigS2}, we provide more examples of the energy spectrum for the generic non-Hermitian model with $\kappa\ge2$. It is not difficult to conclude that the above discussions on the energy spectrum of the minimal model without mosaic modulation also apply to mosaic case with large $\kappa$.
Apparently, energy-dependent localization does occur as the strength of the on-site quasiperiodic potential $\lambda$ varies. However, no evident real-complex transition in energy shows up. Thus, one needs to study the non-Hermitian mobility edge physics for this family of generic non-Hermitian models utilizing the full complex energy domain.

\section{A rough estimation of the range of energy spectrum}\label{RangofE}
Utilizing Avila's global theory, one can analytically calculate the Lyapunov exponent $\Gamma(E)$.
And by letting $\Gamma(E)=0$, a preliminary form of the mobility edge is obtained. Then, its intersection with the system's actual energy spectrum gives the true mobility edge.
Accordingly, one needs to be clear about the actual range of the model's energy spectrum in advance.
Here we present a rough estimation of the range of energy spectrum for the generic non-Hermitian qasiperiodic model in Eq.(\ref{vavbEqa}), which features a on-site potential as follows,
\begin{equation}
		V_{j}=\left\{
		\begin{aligned}
			\frac{\lambda e^{i\gamma}\cos(2\pi j b+\theta)+\delta}{1-\alpha\cos(2\pi j b+\theta)},\quad & j=m\kappa,\\
			0,\quad & otherwise.
		\end{aligned}\right.
	\label{model}
	\end{equation}	
As shown in Sec.~\ref{spectrum}, it is clear that the eigenvalues of model Eq.(\ref{vavbEqa}) are generally complex. Thus naturally, we'd better estimate their range separately for the real parts and the imaginary parts.

According to the operator theory, the range of the physical possible energy spectrum $E$ of the model Eq.(\ref{vavbEqa}) can be estimated as
\begin{equation}
Re(E)\subseteq \left[-2\left|t\right|+min(Re(V_{j})),2\left|t\right|+max(Re(V_{j})) \right],
\label{Rerange}
\end{equation}
and
\begin{equation}
Im(E)\subseteq \left[min(Im(V_{j})),max(Im(V_{j})) \right].
\label{Imrange}
\end{equation}

\textbf{Firstly}, we present in detail the estimation process for the range of the real part $Re(E)$ of the eigenvalue.

For the minimal model with $\kappa=1$, apparently we have,
\begin{align}
Re(V_{j})&=\frac{\lambda\cos(\gamma)\cos(2\pi j b+\theta)+\delta}{1-\alpha\cos(2\pi j b+\theta)}.
\end{align}
Upon this point, we make a symbol notation:
$\Lambda=\lambda\cos(\gamma)/\alpha$.
Then, the real part of the on-site potential for the $\kappa=1$ case can be rewritten as,
\begin{equation}
\begin{aligned}
Re(V_{j})=\frac{\Lambda+\delta}{1-\alpha\cos(2\pi j b+\theta)}-\Lambda
\end{aligned}
\end{equation}

Therefore, for the case with $\kappa=1$, the range of the real part of the eigenvalue can be straightforwardly given as follows.

(i) when $\Lambda+\delta\textgreater0$, we have,
\begin{equation}    		
Re(E) \subseteq\left[-2\left|t\right|+\frac{\Lambda+\delta}{1+\left|\alpha\right|}-\Lambda,2\left|t\right|+\frac{\Lambda+\delta}{1-\left|\alpha\right|}-\Lambda\right].
    		\label{k=1,1}
    	\end{equation}

(ii) when $\Lambda+\delta\textless0$, we have,
\begin{equation}    	Re(E)\subseteq\left[-2\left|t\right|+\frac{\Lambda+\delta}{1-\left|\alpha\right|}-\Lambda,2\left|t\right|+\frac{\Lambda+\delta}{1+\left|\alpha\right|}-\Lambda\right].
\label{k=1,2}
\end{equation}

\begin{figure}[h]
\centering
\includegraphics[width=9.7cm]{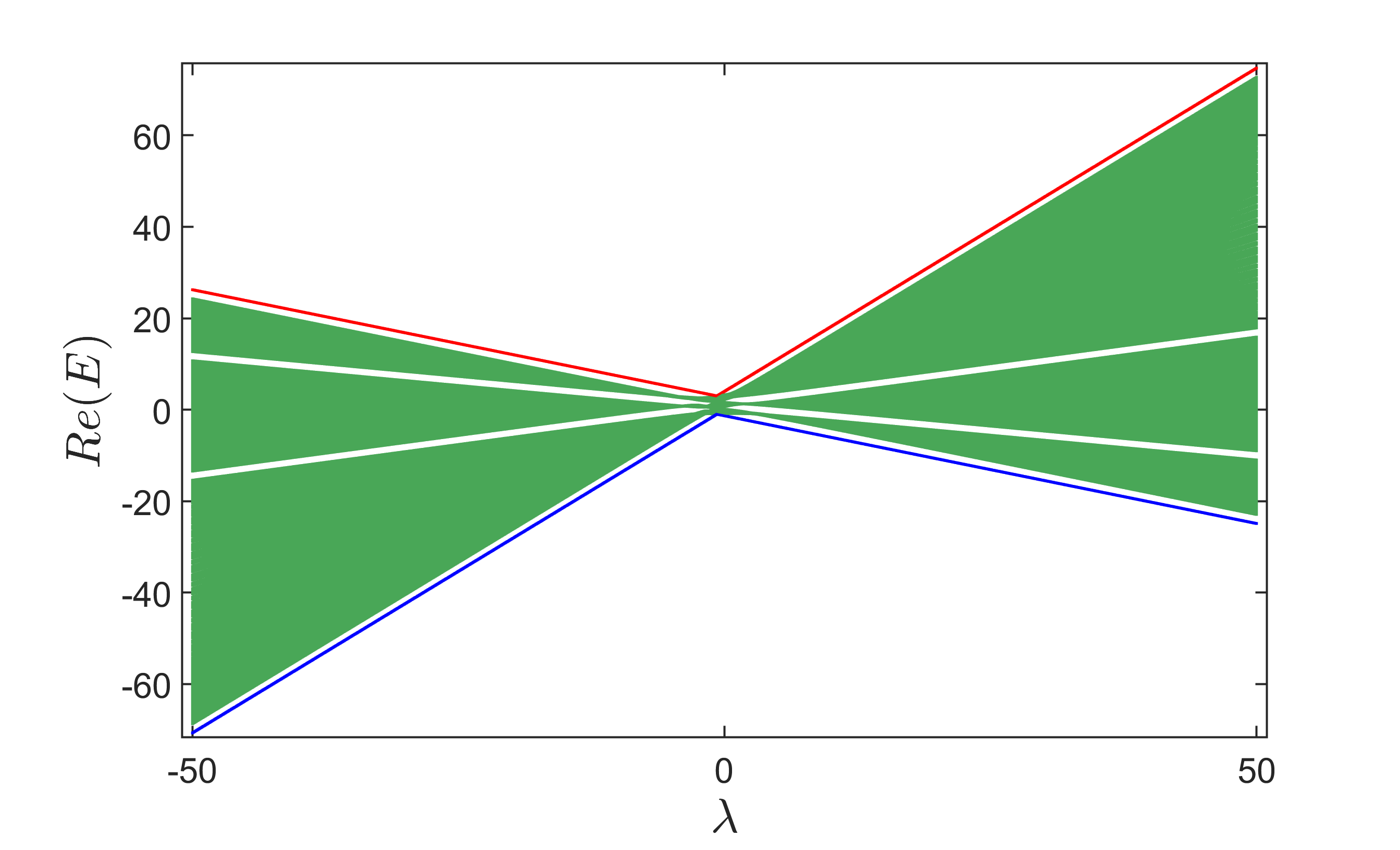}
\caption{ An example to demonstrate the validity of the estimation of the range of the real parts of eigenvalues for the simplest case with $\kappa=1$. The red line denotes the analytically estimated upper bound of the range of $Re(E)$ and the blue line denotes the lower bound. Model parameters in this example are listed as follows,
$L=377$, $t=1$, $\delta=1$, $\theta=0$, $\alpha=0.5$, and $\gamma=\pi/4$.}
\label{K1RE}
\end{figure}

The validity of the above estimation process for $Re(E)$ is exemplified in Fig.\ref{K1RE}.

However, when it comes to cases with $\kappa\ge 2$, one needs to further take into account those lattice sites with zero on-site potential to determine $max(Re(V_{j}))$ and $min(Re(V_{j}))$.
Similar to the simplest case with $\kappa=1$, it can also be divided into two cases for analysis, but now each case is further subdivided into three sub-cases.
Thus we have the following result.

For case (i), $\Lambda+\delta \textgreater 0$. It has the following three sub-cases.

When $\frac{\Lambda+\delta}{1+\left|\alpha\right|}-\Lambda\textgreater0$, the value of $min(Re(V_{j})$ should be $0$. Accordingly, we have,
\begin{equation}    	Re(E)\subseteq\left[-2\left|t\right|,2\left|t\right|+\frac{\Lambda+\delta}{1-\left|\alpha\right|}-\Lambda\right].
\end{equation}

When $\frac{\Lambda+\delta}{1-\left|\alpha\right|}-\Lambda\textless0$, the value of $max(Re(V_{j}))$ should be $0$.
\begin{equation}    	Re(E)\subseteq\left[-2\left|t\right|+\frac{\Lambda+\delta}{1+\left|\alpha\right|}-\Lambda,2\left|t\right|\right].
\end{equation}

When $\frac{\Lambda+\delta}{1+\left|\alpha\right|}-\Lambda\textless 0 \textless \frac{\Lambda+\delta}{1-\left|\alpha\right|}-\Lambda$, the range of $Re(E)$ falls back to the same result as $\kappa=1$ case.
\begin{equation}    	Re(E)\subseteq\left[-2\left|t\right|+\frac{\Lambda+\delta}{1+\left|\alpha\right|}-\Lambda,2\left|t\right|+\frac{\Lambda+\delta}{1-\left|\alpha\right|}-\Lambda\right].
\end{equation}

For case (ii), $\Lambda+\delta \textless 0$. Similarly, it also has three sub-cases as following.

When $\frac{\Lambda+\delta}{1-\left|\alpha\right|}-\Lambda\textgreater0$, the value of $min(Re(V_{j})$ should be $0$. Accordingly, we have,
\begin{equation}    	Re(E)\subseteq\left[-2\left|t\right|,2\left|t\right|+\frac{\Lambda+\delta}{1+\left|\alpha\right|}-\Lambda\right].
\end{equation}

When $\frac{\Lambda+\delta}{1+\left|\alpha\right|}-\Lambda\textless0$, the value of $max(Re(V_{j}))$ should be $0$.
\begin{equation}    	Re(E)\subseteq\left[-2\left|t\right|+\frac{\Lambda+\delta}{1-\left|\alpha\right|}-\Lambda,2\left|t\right|\right].
\end{equation}

When $\frac{\Lambda+\delta}{1-\left|\alpha\right|}-\Lambda\textless 0 \textless \frac{\Lambda+\delta}{1+\left|\alpha\right|}-\Lambda$, the range of $Re(E)$ also falls back to the same result as $\kappa=1$ case.
\begin{equation}    	Re(E)\subseteq\left[-2\left|t\right|+\frac{\Lambda+\delta}{1-\left|\alpha\right|}-\Lambda,2\left|t\right|+\frac{\Lambda+\delta}{1+\left|\alpha\right|}-\Lambda\right].
\end{equation}

\begin{figure}[h]
\centering
\includegraphics[width=16.7cm]{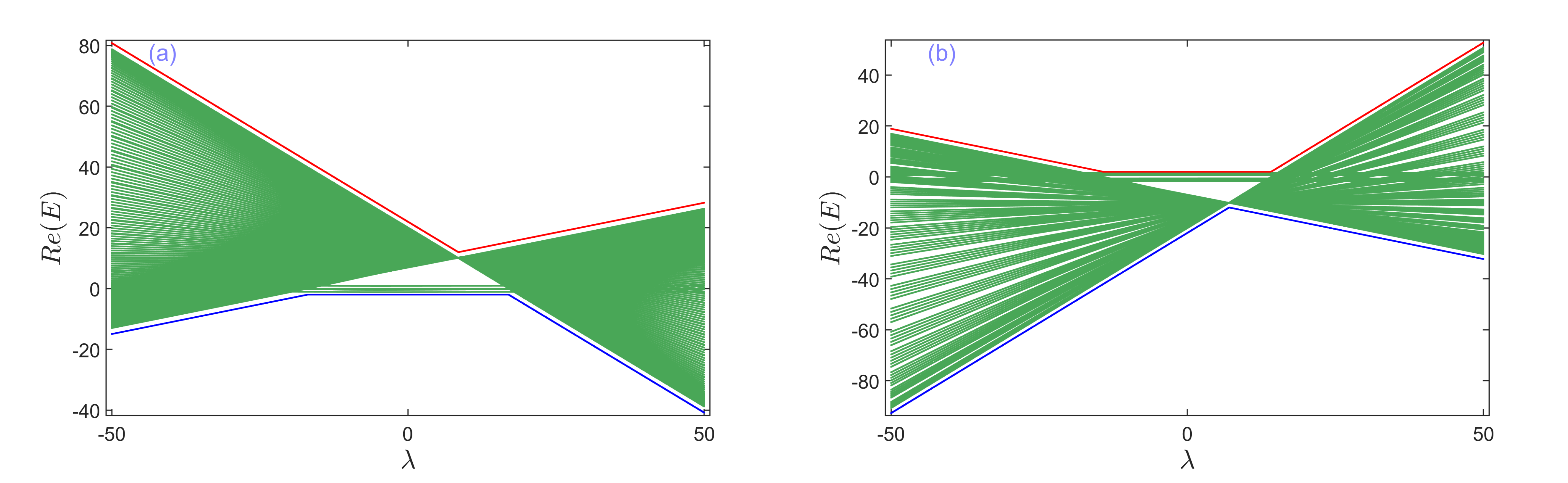}
\caption{
Examples to show the validity of the estimation of the range of the real parts of eigenvalues for mosaic cases with larger $\kappa$s. (a) $\kappa=2$. Other model parameters are as follows, $L=377$, $t=1$, $\delta=10$, $\theta=0$, $\alpha=0.5$, and $\gamma=7\pi/10$.
(b) $\kappa=3$. And other model parameters are as follows, $L=377$, $J=1$, $\delta=-10$, $\theta=0$, $\alpha=0.5$, and $\gamma=\pi/4$.
The red line denotes the analytically estimated upper bound of the range of $Re(E)$ and the blue line denotes the lower bound. }
\label{K2K3RE}
\end{figure}

To demonstrate the validity of the estimation process of the range of the real parts of eigenvalues for mosaic cases with larger $\kappa$s, we numerically provide two examples in Fig.\ref{K2K3RE}.

\textbf{Secondly}, we provide the estimation process for the range of the imaginary part $Im(E)$ of the eigenvalue. The deduction process is much briefer than that of the real part discussed above.

Apparently for any $\kappa$, the imaginary part of the on-site potential is always in a form as,
\begin{align}
Im(V_{j})&=\frac{\lambda\sin(\gamma)\cos(2\pi j b+\theta)}{1-\alpha\cos(2\pi j b+\theta)}.
\end{align}
For ease of representation, we could define $\Delta=\lambda\sin(\gamma)/\alpha$. Then, the imaginary part of the on-site potential can be rewritten as,
\begin{equation}
\begin{aligned}
Im(V_{j})=\frac{\Delta}{1-\alpha\cos(2\pi j b+\theta)}-\Delta.
\end{aligned}
\end{equation}

\begin{figure}[h]
\centering
\includegraphics[width=16.7cm]{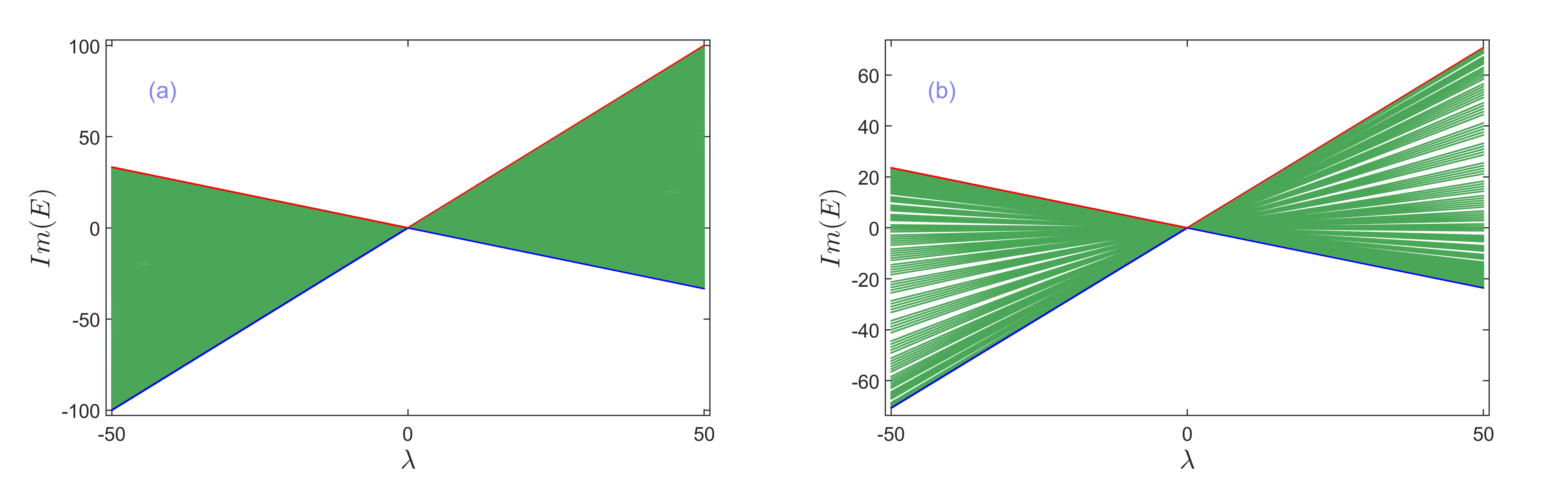}
\caption{Exemplification of the validity of the estimation of the range of the imaginary parts of eigenvalues for the family of generic non-Hermitian quasiperiodic models. (a) $\kappa=1$. Other model parameters are as follows, $L=377$, $t=1$, $\delta=1$, $\theta=0$, $\alpha=0.5$, and $\gamma=\pi/2$.
(b) $\kappa=3$. And other model parameters are as follows, $L=377$, $J=1$, $\delta=-10$, $\theta=0$, $\alpha=0.5$, and $\gamma=\pi/4$.
The red line denotes the analytically estimated upper bound of the range of $Im(E)$ and the blue line denotes the lower bound. }
\label{K1K3IM}
\end{figure}

Therefore, on one hand, for the case $\Delta>0$, we have
\begin{equation}    	Im(E)\subseteq\left[\frac{\Delta}{1+\left|\alpha\right|}-\Delta,\frac{\Delta}{1-\left|\alpha\right|}-\Delta\right],
\end{equation}
while on the other hand, for $\Delta\textless0$, we have
\begin{equation}    	Im(E)\subseteq\left[\frac{\Delta}{1-\left|\alpha\right|}-\Delta,\frac{\Delta}{1+\left|\alpha\right|}-\Delta\right].
\end{equation}
We provide numerical verifications for the analytical estimation of the range of the imaginary parts of eigenvalues for the family of generic non-Hermitian quasiperiodic models in Fig.\ref{K1K3IM}

Accordingly, for the parameter settings corresponding to Fig.\ref{Fig01}(b) in the main text, the range of the imaginary part $Im(E)$ of the model's actual spectrum is easily obtained as $[-2/3,2]$. The intersection of the range $[-2i/3,2i]$ in complex plane with the exact non-Hermitian mobility edge given by Eq.(\ref{k1ME}) produce the true mobility edge which is shown in Fig.\ref{Fig01}(b) in the main text.

\begin{figure}[h]
\centering
\includegraphics[width=9.7cm]{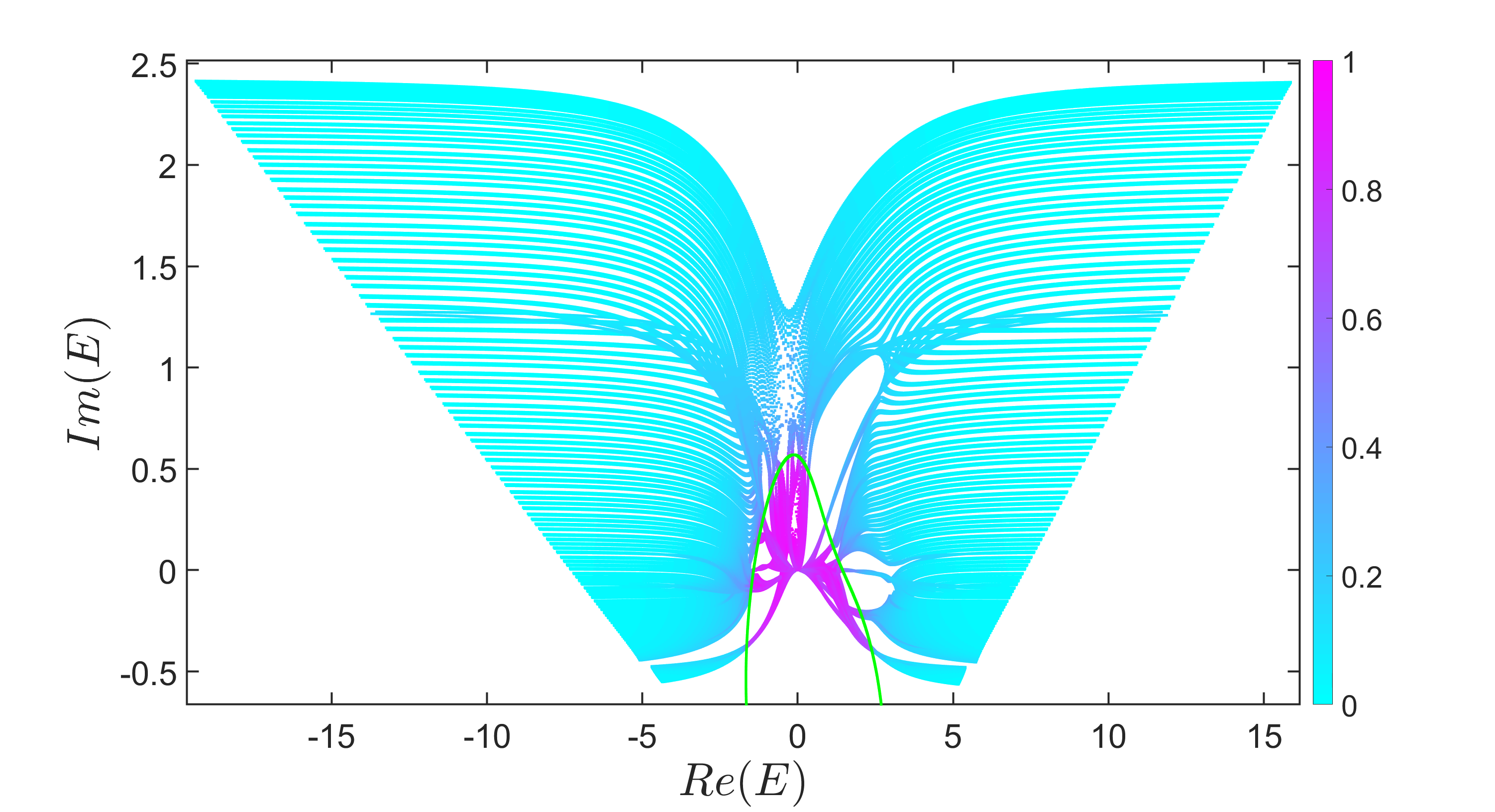}
\caption{
A non-Hermitian butterfly spectrum example for the generic non-Hermitian mosaic quasiperiodic model with $\kappa=2$. Fractal dimension (FD) of each eigenstate is denoted by the color of each energy point in the spectrum. The exact non-Hermitian mobility edge is denoted by a green line which is obtained by numerically solving the exact relation Eq.(\ref{MErelation}) and considering the actual range of the model's spectrum at the same time.
Parameters: $L=610$, $\lambda=1.2$, $\alpha=0.6$, $\gamma=0.7\pi$, $\theta=0$, $t=1$, and the modulation parameter $\delta$ varies from $-7$ to $7$.}
\label{K2ME}
\end{figure}
\begin{figure}[h]
\centering
\includegraphics[width=9.7cm]{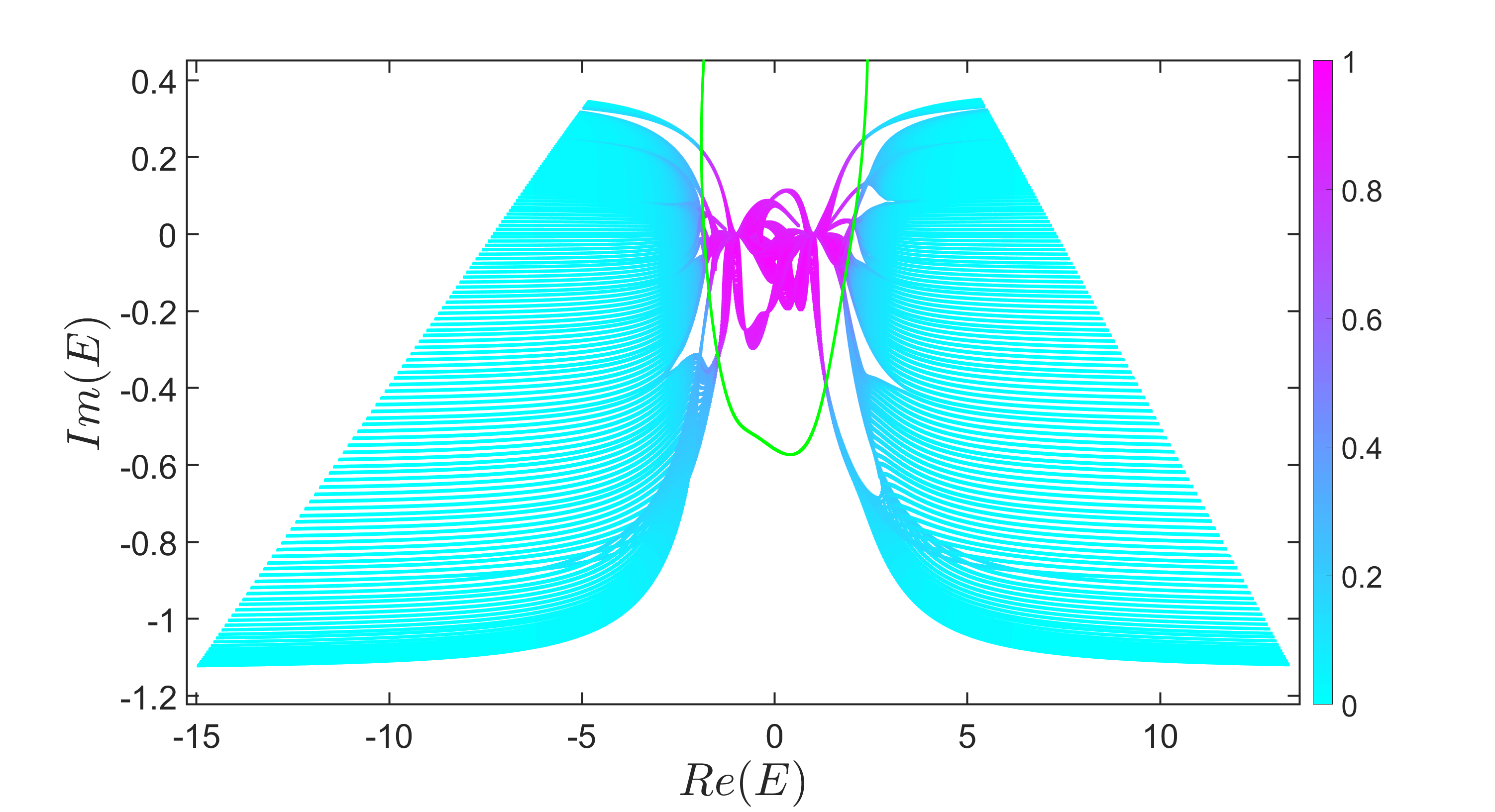}
\caption{A non-Hermitian butterfly spectrum example for the generic non-Hermitian mosaic quasiperiodic model with $\kappa=3$. Fractal dimension (FD) of each eigenstate is denoted by the color of each energy point in the spectrum. The exact non-Hermitian mobility edge is denoted by a green line which is obtained by numerically solving the exact relation Eq.(\ref{MErelation}) and considering the actual range of the model's spectrum at the same time.
Parameters: $L=987$, $\lambda=0.7$, $\alpha=-0.5$, $\gamma=0.3\pi$, $\theta=0$, $t=1$, and the modulation parameter $\delta$ varies from $-7$ to $7$.}
\label{K3ME}
\end{figure}

\section{The non-Hermitian Butterfly Spectra and the exact NHMEs for cases with larger $\kappa$s}\label{largeK}
In the main text, for simplicity and clarity, we mainly focus on the minimal model without mosaic modulation and show the exact non-Hermitian mobility edge along with the exotic butterfly spectrum in the full complex plane. The analytical derivation of the exact non-Hermitian mobility edge is provided in the main text, where we mentioned that the analytical deduction process also applies to general mosaic cases with larger $\kappa$s. In this section, we provide some examples for the non-Hermitian butterfly spectra and the exact NHMEs for generic non-Hermitian quasiperioic mosaic models with larger $\kappa$s.
Although it is usually difficult to obtain a compact formula for the exact NHME of the model with large $\kappa$, one can resort to numerical calculations to obtain the exact result of the NHME. As concrete examples, we present more details for two mosaic cases with $\kappa=2$ and $\kappa=3$ in the following.

In the main text, by using Avila's global theory, we have analytically obtained the Lyapunov exponent for the model in the general case as
\begin{align}
\Gamma(E)=\frac{1}{\kappa}\mathrm{max}\{\ln  \frac{2 f}{1+\sqrt{1-\alpha^2}} , 0\},
\end{align}
in which,
\begin{align}
f=\mathrm{max}\left\{ \left |\frac{2\alpha a_{\kappa -1}-\chi a_{\kappa}\pm \sqrt{\chi^2 a_{\kappa}^2-4\alpha\chi a_{\kappa}a_{\kappa-1}+4\alpha^2 a_{\kappa} a_{\kappa-2}}}{4} \right | \right \}.
\end{align}
Then NHMEs can be determined exactly by letting $\Gamma(E)=0$. Therefore, an exact relation is obtained as follows,
\begin{align}
\frac{2 f}{1+\sqrt{1-\alpha^2}}=1,
\label{MErelation}
\end{align}
which basically produces the exact non-Hermitian mobility edge.
Simplifying Eq.(\ref{MErelation}), an elegant and compact analytical formula could be obtained for the case $\kappa=1$. However, for cases with $\kappa \ge 2$, such a analytical formula for NHME is generally either very long and complex or difficult to obtain. Therefore, we mainly adopt numerical methods for the later cases instead. As concrete examples, in Fig.~\ref{K2K3ME}, Fig.\ref{K2ME} and Fig.\ref{K3ME} we give out the exact NHMEs for the cases with $\kappa=2$ and $\kappa=3$ by solving the exact relation Eq.(\ref{MErelation}) numerically.

\end{document}